\documentclass[conference]{IEEEtran}
\IEEEoverridecommandlockouts
\usepackage{cite}
\usepackage{amsmath,amssymb,amsfonts}
\usepackage{algorithmic}
\usepackage{graphicx}
\usepackage{textcomp}
\usepackage{xcolor}
\usepackage{epsfig}
\usepackage{url}
\usepackage{verbatim}
\usepackage[ruled,vlined]{algorithm2e}

\def\BibTeX{{\rm B\kern-.05em{\sc i\kern-.025em b}\kern-.08em
    T\kern-.1667em\lower.7ex\hbox{E}\kern-.125emX}}
\begin{document}

\title{Privacy-preserving Identity Broadcast for  Contact Tracing Applications}

\author{
\IEEEauthorblockN{Vladimir Dyo}
\IEEEauthorblockA{\textit{School of Computer Science and Technology} \\
\textit{University of Bedfordshire, UK}\\
vladimir.dyo@beds.ac.uk}
\and
\IEEEauthorblockN{Jahangir Ali}
\IEEEauthorblockA{\textit{School of Computer Science and Technology} \\
\textit{University of Bedfordshire, UK}\\
jahangir.ali@study.beds.ac.uk}
}

\IEEEoverridecommandlockouts
\IEEEpubidadjcol
\maketitle

\begin{abstract}
Wireless Contact tracing has emerged as an important tool for managing the COVID19 pandemic and relies on continuous broadcasting of a person's presence using Bluetooth Low Energy beacons. 
The limitation of current contact tracing systems in that a reception of a $single$ beacon is sufficient to reveal the user identity, potentially exposing users to malicious trackers  installed along the roads, passageways, and other infrastructure. 
In this paper, we propose a method  based on Shamir secret sharing  algorithm,  which lets mobile nodes reveal their identity  only after a certain predefined contact duration, remaining invisible to  trackers with  short or fleeting encounters. 
Through data-driven evaluation, using a dataset containing 18 million BLE sightings, we show that the method  drastically reduces the privacy exposure of users. 
Finally, we implemented the approach on Android phones to demonstrate its feasibility  and measure performance for various network densities.  
\end{abstract}


\maketitle
\section{Introduction}

\IEEEPARstart{C}{ontact} 
 tracing has emerged as an essential tool for breaking the chain of coronavirus infections and containing the COVID-19 pandemics. 
While many countries around the world have been adopting contact tracing apps, user exposure to continuous tracking have raised numerous privacy concerns \cite{gvili2020contact}. 
As per current Center of Disease Control (CDC) guidelines \cite{cdc}, the close contact is defined as a cumulative total of 15 minutes within 6 feet within 24 hours, however  many contact tracing applications are able to identify user identity from a single beacon sighting, which  exposes a user to unwanted tracking. 
The potential loss of privacy can discourage users from using the apps for fear of abuse or discrimination. 

In response to the pandemic, Google and Apple developed Bluetooth-based contact tracing protocol based on DP-3T \cite{dp3t}. 
The protocol provides location privacy by randomising identifiers using pseudorandom number generator seeded by daily key that is kept secret unless the user decides to share it in case of infection.
Once the daily key is shared, it is used by other users to recreate the same sequence of IDs to check if and when they have been in contact with the infected person. 
However, the approach has two well-documented limitations. 
Firstly, revealing the daily key allows other users to reconstruct all daily identifiers for that day, and therefore
exposes the infected user to tracking throughout an entire day. 
This could potentially lead to revealing infected users' daily trajectory, a location of his or her home or office \cite{ali2021icoin}.
Secondly, the identity of an infected user becomes known not only to people who have been in direct continuous contact with the infected person, but also to  Bluetooth devices that only  had a fleeting encounter with the person. 

\begin{figure}[h]
\centering
\includegraphics[width=2in]{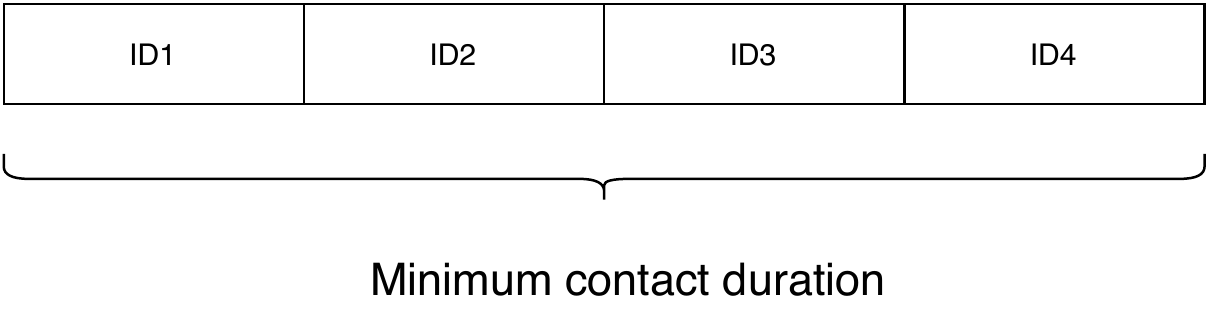}
\caption{Cross-hashing enables defining a minimum contact duration.}
\label{fig:crosshashing}
\end{figure}

In our recent work, \cite{ali2021icoin} we  proposed a cross-hashing method, where infected users share a hash computed from a chain of sequential randomised beacons, rather than their daily keys.
This effectively enforces the minimum contact duration as nearby users need to stay around for long enough to receive and compute a hash chain from subsequent beacons. 
However, in a noisy wireless environment, where beacons can be dropped or corrupted due to interference or collisions, the cross-hashing approach  requires defining and sharing multiple versions of the same chain as shown in Fig. \ref{fig:crosshashing}.
In other words, an infected user may have to compute and share multiple combinations of subsequent beacons to account for lost or corrupted beacons. 
For example, to account for a single packet failure, a user needs to compute and share multiple versions of the same hash chain: $H(id1,id2,id3)$, $H(id2,id3,id4)$, $H(id1,id3,id4)$ and $H(id1,id2,id4)$ for the same contact event, Fig. \ref{fig:crosshashing}. 
This creates additional communication overhead, especially for high frequent beacons and complicates the implementation. 
\vspace{-2mm}

In this paper we propose an alternative solution that enforces minimum contact duration while taking into account an unreliable nature of wireless channel. 
The proposed method is based on using Shamir's secret share algorithm, which divides the device identity into $n$ parts, such that any $k< n$ parts are required to reconstruct the device identifier.
The approach provides protection against opportunistic scanners as receiving  fewer number than $k$ of beacons makes it impossible  to reconstruct the mobile device identity. 
We evaluated the approach using Bluetooth Low Energy traces and showed that it can drastically reduce the amount of spatio-temporal exposure.
In particular, the approach reduces the number of registered encounters from 18,655,641 to just 103,129 when using ($k$=5, $n$=6) scheme.  
Finally, we show that the approach can be  implemented  on Android operating system for proximity tracking applications. 
To the best of our knowledge, this is the first work that proposes splitting user identifiers to enforce minimum contact duration and reduce  users' exposure to   opportunistic tracking. 

\section{Related work}
The Bluetooth Low Energy (BLE) contact tracing applications rely on short-range Bluetooth transmissions to advertise presence and detect nearby devices. 
To protect users against tracking, BLE defines private addresses which are randomly generated and rotated periodically within the payload.
Therefore, the contact tracing apps use encrypted user identifier to share information about the person's presence.  

DP-3T based systems \cite{dp3t}\cite{gaen}, broadcast  a random identifier  generated  using pseudorandom function keyed by a daily Temporal Exposure Key (TEK). 
This identifier, called  Rolling Proximity Identifier (RPI),  is generated every 5 minutes resulting in 48 unique identifiers per day. 
In case a user tests positive for coronavirus, his recent Temporal Exposure Keys  are shared with other users, so that they can re-generate  identifiers, compare with  their logged data and determine whether they have been in contact with an infected user. 
However, by doing so, the users will also be able to reconstruct the entire daily trajectory of the infected user. 
The problem has been addressed in our recent work \cite{ali2021icoin}, where we proposed a cross-hashing approach, in which the daily temporal key is kept secret at all times. Instead, in case of infection, a user reveals cross-contact identifiers (CCI), which are obtained by hashing sequential randomised identifiers. 
Sharing CCI limits the privacy exposure to a single encounter only. Secondly, obtaining CCI requires hashing sequential identifiers, and therefore, enforces minimum contact duration for revealing the identity. 

BLE-based contact tracing apps use either connection-oriented or connection-less communication models. 
In connection-oriented model, the devices advertise their contact tracing capability by including a corresponding service UUID in the advertising packets. 
The devices then  establish a connection with each other to track proximity and exchange data. 
In this mode, since the connection is established using private address, the device can reveal user identifier after a certain duration of time. 
This model is used by UK's NHS Covid App \cite{nhscovid}, BlueTrace-based systems \cite{bluetrace}.
However, the connection-oriented model may have scalability issues, as device needs to establish connection with each nearby device. 
Since many BLE implementations limit the maximum number of concurrent connections, the model can potentially be vulnerable to denial of service attacks, where a malicious user establishes random connections preventing a user from connecting to genuine users.  
Furthermore, to maintain connection, the BLE identifier needs to remain the constant, which in case of long encounters can compromise user identity. 

In connection-less model, also known as a broadcast model, devices periodically broadcast advertising packets to advertise device's presence and  collect information about other devices. 
The model is inherently scalable and simple to implement as it requires no interaction between devices apart from broadcasting beacons and logging beacons from other devices. 
The connection-less model is used in DP-3T  based systems \cite{dp3t}.
The key limitation of model, in our view, is that transmission of a single identifier can reveal a user identifier. 
Finally, hybrid models support both approaches depending on device capability and OS limitations. 
For more detailed analysis and comparison of both connection models, we refer the reader to  \cite{cunche2020insa}. 
Finally, Cunche {\em et al} \cite{cunche2020insa}  discuss an idea of splitting the advertising payload into multiple blocks to be advertised one after another in separate advertisement packets.
However, this was discussed as a method to cope with large payloads, rather than reducing the privacy exposure of users.  

\cite{pact}\cite{pact2}\cite{hatke2020using} describe an approach on optimising detection performance for automated coronavirus contact tracing applications. 
In particular, PACT\cite{pact}\cite{pact2} proposes a method that makes a binary decision based on received BLE signal strength, on whether the two phones have been "Too Close For Too Long" (TCFTL). 

In this paper we present an alternative solution, where a device identity is split into multiple shares. 
This allows neighbour nodes to reconstruct the identifier after spending sufficient time in the proximity of the mobile device.  
This also does not require a centralised system to distribute the CCIs, and can be more suitable for decentralised implementation.

\section{Shamir's shared secret algorithm}

Shamir's secret sharing  is a cryptographic algorithm, where a secret is divided into $n$ parts, such that any $k < n$ parts are required to reconstruct an original key.  
The minimum number of parts, called shares, $k$ needed to reconstruct the secret is called a threshold. 
Importantly, the knowledge of  shares fewer than $k-1$ is not sufficient to reconstruct the whole key. 
The scheme is known as $(k, n)$ threshold scheme. 
If $k = n$ then reconstructing the key requires all shares. 
This is a trivial case, where a secret is simply split into $n$ parts. 
The algorithm is used in high security environments for key distribution among multiple individuals \cite{Shamir1979acm}. 
It is also used in bitcoin's multisig feature, where a transaction is approved by multiple signatories \cite{Boneh2018aicr}. 

The geometric intuition behind the scheme is  that $k$ points are sufficient to define an $(k-1)$ degree polynomial. 
For example, two points are sufficient to define a line, 3 points to define a parabola and so on. 
To construct a $(k,n)$ scheme for sharing a secret $s$, the scheme creates the polynomial 
$f(x)= s +a_{1}x+a_{2}x^{2}+a_{3}x^{3}+\cdots +a_{k-1}x^{k-1}$, where $a_1, a_2,...a_{k-1}$ are chosen randomly, uniformly and independently.  
The scheme then computes $y_i = f(i)$, for all $1\leq i\leq n$. 
Each share is given a point, which consists of an input to the polynomial,  the corresponding  output and  the prime number which defines the finite field to use. 
The secret $s$ can be computed using Lagrange interpolating formula, given any subset of $k$ of these pairs:

\begin{equation}
s = f(0) = \sum_{l = 1}^{k} y_{j} \displaystyle\prod_{v = 1,v \neq l}^{k}\frac{-x_v}{x_l - x_v} mod \, p 
\end{equation}

\section{Contact duration based Identity-disclosure}

\begin{figure}[h]
\center
\includegraphics[width=8.2cm]{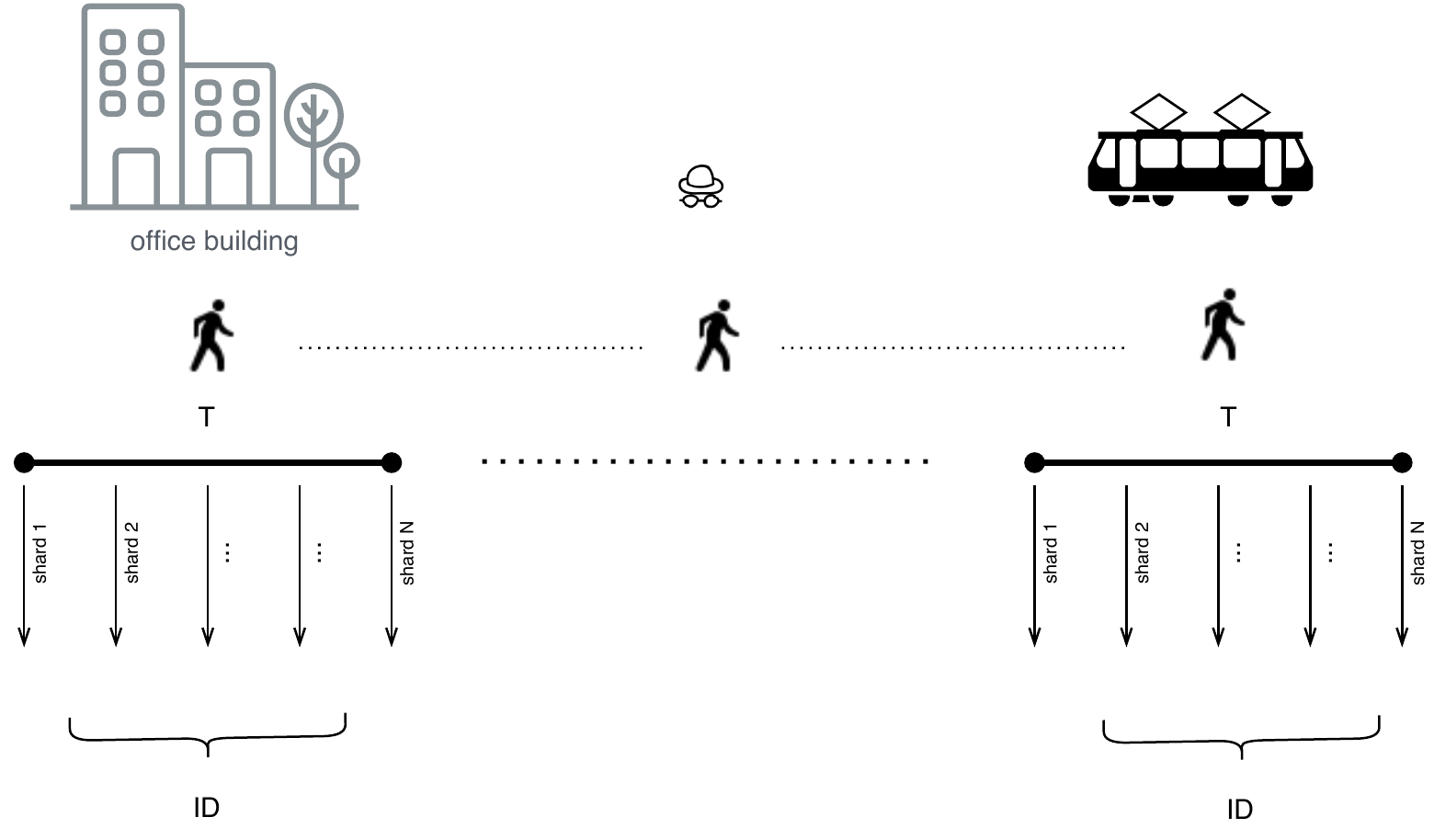}
\caption{Motivating scenario. In places where a person spends sufficient time $T$, such as office or public transport, neighbour nodes are able to collect $K < N$ shares (shards) to reconstruct the node's ID, while the scanners and random passerby  along the road are not able to track the user.\label{fig:scenario}}
\end{figure}

A motivating scenario is presented in Fig. \ref{fig:scenario}. 
In places, where a person spends  {\em sufficient amount of time} such as office or within public transport, his or her ID become visible to other users. 
The Bluetooth scanners that have a fleeting contact with the user, should not be able to see user's identity. 
Intuitively, this  behaviour is similar to  human interactions in real life in that we do not reveal our identity immediately but only after a certain time required to establish trust. 
Similarly, the proposed approach requires any scanner to spend a certain minimum duration  near a mobile device to decode its identity as described in the following subsections. 

\subsection{Generating shares}
A device identifier is a 16-byte string, which consist of 12-byte string generated randomly at the initialisation stage and its  CRC32 checksum. 
The  device identifier is split into $n$ shares using Shamir's shared secret ($k$, $n$) scheme with each share having a length of 16-bytes and a 1-byte share ID  numbered from 1 to $n$. 
Reconstructing a device identifier requires at least $k$ shares together with their related share IDs. 
The shares expire after a certain timeout, after which they are regenerated from the same device identifier. 
The expiry is needed to force scanners to keep reconstructing device identifier from new shares to know about device's presence.

\subsection{Transmitting shares}
Each share is transmitted for duration of time $T_{share}$, after which a mobile device  picks another random share to transmit. 
After all $n$ shares have been transmitted, a mobile device generates a new set of shares.  
Thus, any nearby scanner  needs to keep reconstructing the shares to be able to see the real identity of neighbour nodes. 

\subsubsection{Bluetooth Low Energy and AltBeacon}
\begin{figure}[!h]
\centering
\includegraphics[width=3.2in]{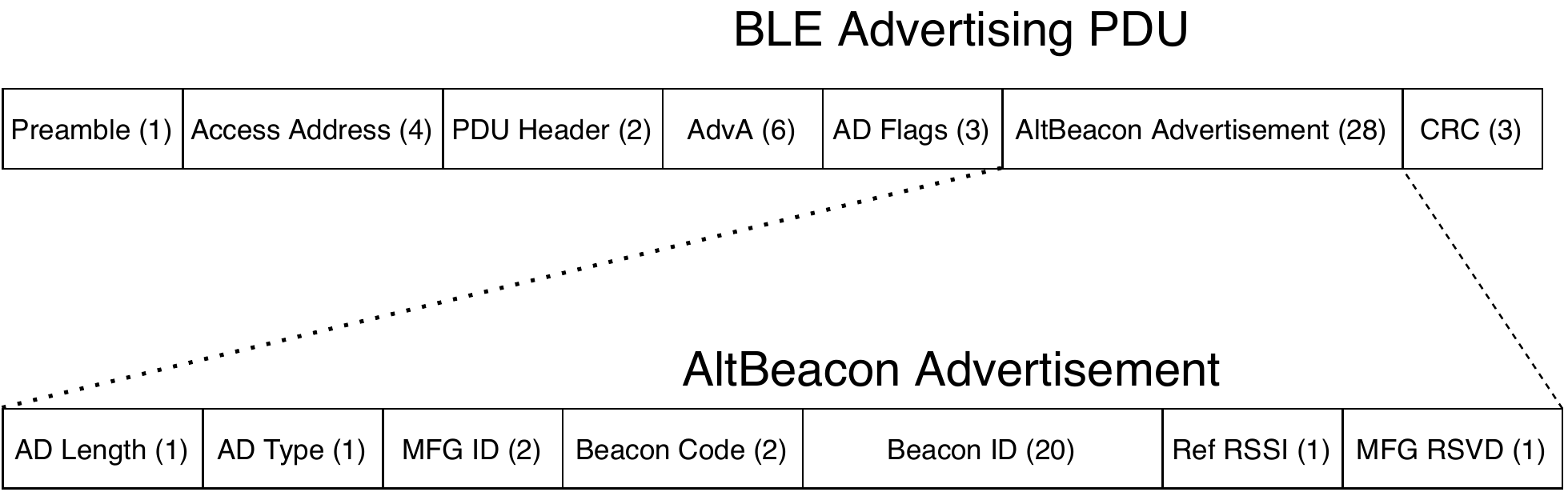}
\caption{BLE and AltBeacon formats \cite{altbeacon}}
\label{fig:altbeacon}
\end{figure}

A secret share and the shareID are carried within  BLE AltBeacon's   Beacon Identifier field. 
AltBeacon \cite{altbeacon} is relatively new format, which has open specification, vendor independent and enjoys library availability for both Android and iOS. 
The format and the fields of the AltBeacon packet is presented in Fig. \ref{fig:altbeacon}. 

Bluetooth Low Energy is part of Bluetooth introduced in 2010, which operates over 2.4\,GHz band over 40 channels, 3 of which are used for adverisements. 
The purpose of advertisements is to advertise the device presence to nearby devices. 
BLE Beacons is one application, where BLE advertisements are used for geolocation purposes. 
There are various BLE standards including Apple's iBeacon \cite{ibeacon} and Google's EddyStone \cite{eddystone}  and an open source AltBeacon format, which was used in our work.

According to the BLE specification, the interval between beacons, $AdvInterval$ should be an integer of 0.625\,ms in the range of 20\,ms to 10.24\,s. 
The BLE 5.2 \cite{ble5} specification extends the maximum value of $AdvInterval$ to 10485\,s. 
To receive an advertisement packet, a device needs to be in a receiving mode, which is characterised by two parameters, $Scan Interval$ and $Scan Window$. 
The former defines the interval between successive scans and the latter defines the duration of each scan. 
Android supports continuous, balanced and low power scan modes. 
In continuous mode, Android scans continuously and discovers nearby devices quickly, however requires high energy consumption and is not suitable for long scans. 
In a low power mode a phone scans for 0.5\,s every 5\,s interval, whereas in balanced it scans for 2\,s every 3\,s interval.

\subsubsection{BLE privacy support}
In addition to having a public address, which is globally unique and fixed, BLE devices can have private addresses, which can be either random static or random private. 
Random static addresses can be either  assigned and fixed or changed during device bootup. Random private addresses can be resolvable or non-resolvable. 
The resolvable random address uses a randomly generated address, which according to specification is changed every 15 minutes. 
The address consists of a pseudorandom number and its hash value computed using Identity Resolving Key (IRK), which is shared with trusted or bonded BLE devices. 
Finally, a non-resolvable random private address is changed periodically but not resolvable by other devices. 
Connection-less contact tracing systems may therefore use non-resolvable random private addresses to keep MAC address private. 

\subsection{Reconstructing identifiers}
\begin{algorithm}[!t]
\footnotesize
\While{true}{
    \If{$tries > maxTries$}{
    break\;
    }
    uniqueShardIDs = workingSetShards.getIDs()\;
    \If{(workingSet.size()$<$k) or (uniqueShardIDs.size()$<$k)}{
        // nothing to reconstruct\;
        break\;
        }
    \eIf{(currentCandidateSet.size() == k)}{
    recovered = ShamirRecover(curentCandidateSet)\;
    \eIf{(crc32(recovered) == beacon.crc32)}{
        // successful reconstruction\;
        remove(currentCandidateSet, workingSet)\;
        currentCandidateSet = \{\}\;
        currentPositionSet = \{\}\;
        break\;
        }{
    // recovered code is illegal, clear attempt\;
    currentPositionSet = \{\}\;
    currentCompositionSet= \{\}\;
    }
    }{
    // add random shard to current composition\;
    sh = getRandomShard(workingSet)\;
    \If{((!currentPositionSet.contains(sh.shardID())) \&\& (!currentCandidateSet.contains(sh.shard())))}{
    // add shard to current composition\;
    currentPositionSet.add(sh.shardID())\;
    currentCandidateSet.add(sh.shard())\;
    }
    }
}
\caption{Simultaneous Identifier reconstruction}
\label{algo}
\end{algorithm}
In the presence of $M$ nodes in the neighbourhood, a scanner  needs to simultaneously reconstruct $M$ identifiers from $M\times n\times k$ shares. 
The reconstruction starts as soon as the scanner accumulates $k$ shares with unique share IDs and is done by probabilistically trying share combination with complementary share IDs. 
Upon reconstruction, a node computes the CRC32 of the recovered identifier and if it matches the CRC32 within the beacon, the reconstruction is considered to be valid. 
Attempting to recover an identifier from  shares that belong to different devices will not result an error but will produce a spurious identifier that will be detected through a checksum. 
The number of combinations required to reconstruct $M$ identifiers from $M \times  n \times k$ shares is equivalent to the number of combinations to pick $M$ sets of $k$ items from a set of $nMk$:
\begin{equation}
Q=\sum_{j=0}^{M-1} \frac{Mn-kj}{Mn-kn}(M-j)^{k-1}
\end{equation}
The search space is reduced significantly due to a shareID, which  allows a scanner  to narrow down share selection to those with  complementing identifiers. 
As an  example, for ($k$=3, $n$=4) scheme, the complementing identifiers would be \{1, 2, 3\}, \{2, 3, 4\} and \{1, 3, 4\}.

Algorithm \ref{algo} shows a pseudocode implementation for simultaneous identifier reconstruction. 
The algorithm is run every time a node receives a new share. 
Once, an identifier is successfully reconstructed, the corresponding shares are linked to the device identifier and are excluded from the further reconstructions. 
As will be shown  in Section \ref{secEvaluation}, an implementation running  ($k$=4, $n$=5) scheme requires on average 56 attempts  to simultaneously reconstruct identifiers from 5 neighbour nodes.

\section{Evaluation}
\label{secEvaluation}

\subsection{Spatio-temporal exposure reduction}
The purpose of evaluation was to show the reduction in total spatio-temporal exposure achieved by Shamir's scheme compared to raw beaconing.
\subsubsection{Data set \& metrics}
The evaluation has been done using BLEBeacon dataset \cite{sikeridis2018blebeacon} containing over 18 million beacon sightings collected from 
from  32 static Raspberry Pi 3 scanners for over a month. 
The scanners  detected the presence of  Bluetooth Low Energy \cite{ble} (BLE)  beacons carried by 99 people who were given Gimbal Series 10 iBeacons  configured to continuously broadcast at 1\,Hz  with a transmission power of 0\,dBm. 
For each detected beacon, a dataset contains a timestamp, beacon ID, scanner ID (edge device), and the RSSI. 
More details about the experiment setup and the dataset  is available at \cite{sikeridis2018blebeacon}.

\begin{figure}[]
\centering
\includegraphics[width=2.6in]{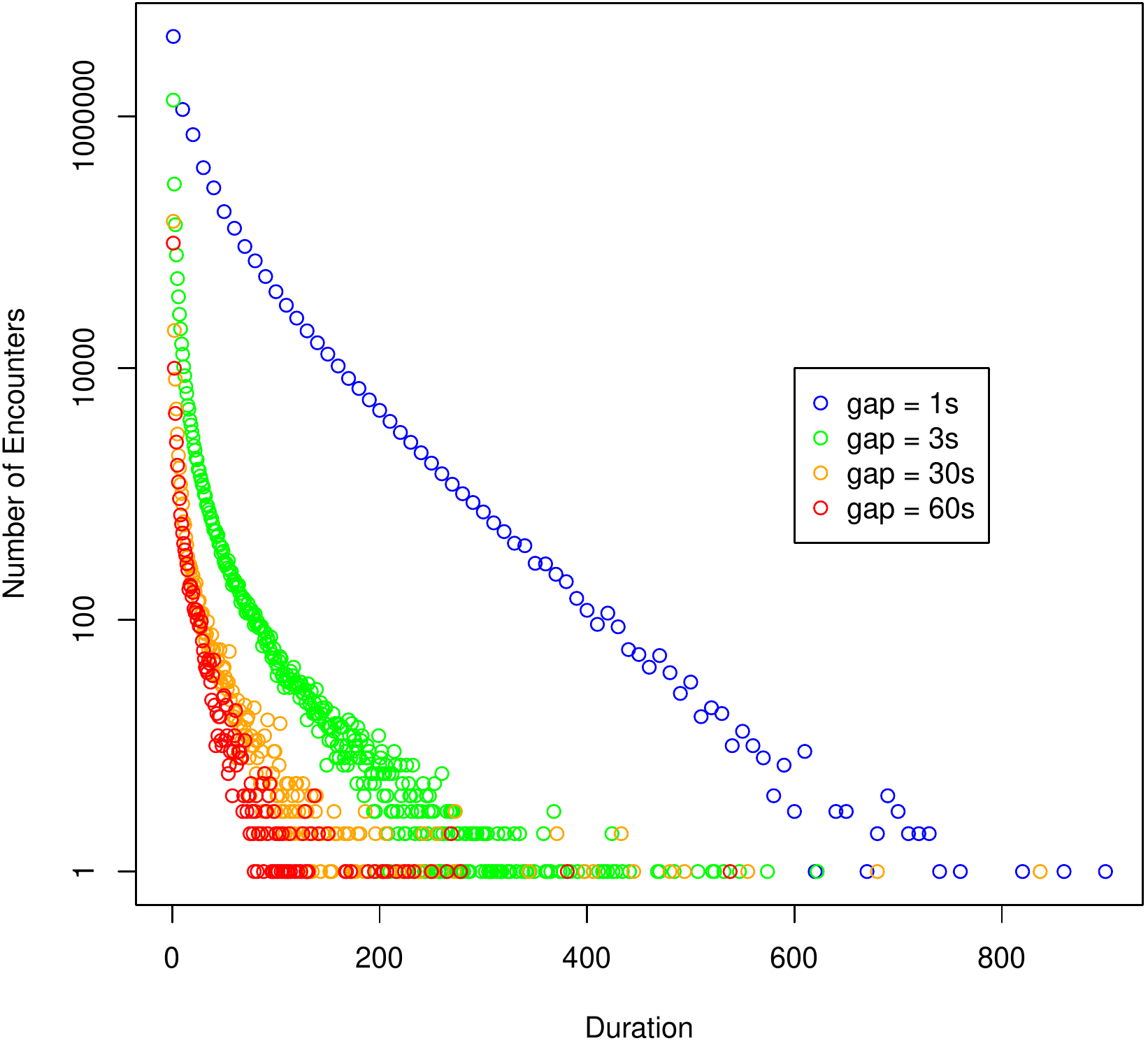}
\vspace{-2mm}
\caption{Distribution of encounter duration vs gap size}
\label{fig:resEncounterStat}
\end{figure}

Fig. \ref{fig:resEncounterStat} shows the distribution of all encounter durations. 
An encounter  is defined as continuous presence with a maximum gap of X=\{1, 3, 30, 60\} seconds between sightings. 
The graph indicates a significant proportion of encounters are very short in duration, often consisting of a single beacon sighting only.  
These encounters, especially single beacon ones, could be effectively filtered out by our approach. 
This behaviour is consistent for all encounter configurations with some differences. 
As   maximum gap increases from 1\,s to 60\,s, the encounters reduce in numbers but also get longer, as several adjacent sightings are more  likely to become merged within a single encounter. 
Still, for all four encounter configurations, the short encounters dominate. 
It should also be noted that the total contact duration, computed as the sum of all encounter durations, increases naturally as smaller gaps become   part of  longer encounters, as shown in Table \ref{tab:encounterStats}.  
An individual's spatio-temporal exposure to a scanner is computed as the total contact duration with the scanner, during which the identity is exposed. 
The total exposure is then computed as a sum of exposures  across all individual and BLE scanner combinations. 
For raw beacons, we assume a single beacon reception exposes a user for 1\,s given the broadcast frequency of 1\,Hz. 
For the proposed approach, the user is considered exposed if a scanner receives $k$ out of $n$ beacons, each containing a share encoded with $(k, n)$ scheme. 
\vspace{-4mm}
\begin{table}[!h]
\tiny
\caption{Encounter statistics.}
\begin{center}
\begin{tabular}{ |c | c | c|  } 
 \hline
  max gap, s  		&  	encounters & total duration 	\\ 
 \hline
1	 	& 	7547457&  18653103	\\ 
 3	 	& 	2108596& 25210586		\\ 
 30	 	&	195955& 36593338	 \\ 
 60	 	&	125196& 39599473	 \\ 
 \hline
\end{tabular}
\end{center}
\label{tab:encounterStats}
\end{table}
\subsubsection{Results}
Table \ref{tab:Simulation}, compares the total spatio-temporal user exposure for various ($k$, $n$) configurations. 
It can be seen that as a threshold of received beacons increases, the  exposure reduces significantly. 
The total raw exposure is shown as the first line in the table ($t$=1, $k$=1, $n$=1) is 18,655,641 seconds. 
For $t$=1, $k$=3, $n$=6 configuration results in 6.8 times exposure reduction and increasing the threshold to $t$=1, $k$=5, $n$=6 configuration reduces the exposure by 180 times. 
This reduction is due to filtering out fleeting encounters where scanners are not able to see the mobile device identities due to short contact time. 
This configuration assumes beaconing at 1\,Hz frequency combined with continuous scanning. 
\begin{table}[h]
\tiny
\caption{Spatio-temporal exposure reduction.}
\begin{center}
\begin{tabular}{ |c|c|c| } 
 \hline
 $t$, $k$, $n$  		&  	slots exposed & total exposure, s \\ 
 \hline
 t=1, (1, 1)	 		& 	18,655,641 	& 18,655,641  \\ 
 t=1, (3, 6)	 		& 	 2,739,950		&  2,739,950\\ 
 t=1, (4, 6)	 		& 	 967,812		&  967,812\\ 
 t=1, (5, 6)	 		& 	 103,129		& 103,129\\ 
\hline
 t=5, (3, 6)	 		& 	1,162,913 	& 5,814,565  \\ 
 t=5, (4, 6)	 		& 	1,035,200 	&  5,176,000\\ 
 t=5, (5, 6)	 		& 	806,185 	& 4,030,925 \\ 
\hline
 t=10, (3, 6) 		 & 	662,295	& 6,622,950 \\ 
 t=10, (4, 6)		 & 	612,878	& 6,128,780  \\ 
 t=10, (5, 6) 	 	& 	537,860	& 5,378,600  \\ 
\hline
 t=20, (3, 6)		& 	379,854	& 7,597,080 \\ 
 t=20, (4, 6) 		& 	355,480	& 7,109,600 \\ 
 t=20, (5, 6) 	 	& 	323,788	& 6,475,760  \\ 
 \hline
\end{tabular}
\end{center}
\label{tab:Simulation}
\end{table}
The second, third and fourth sections   in the table show the effect of a more practical configuration, where a scanner needs to receive $k$ out of $n$ shares, when each share is transmitted for $t$ seconds (i.e. $t$ times). 
In particular, the second configuration corresponds to  1\,Hz beacon frequency with an Android low power mode scanning. 
As can be seen $t$=5, $k$=5, $n$=6 configuration results in 3.47 times exposure reduction,  which 
further reduces to 2.88 for $t$ = 20.

\subsection{Simultaneous reconstruction performance}
Table \ref{tab:simulReconstruction} shows the performance results for simultaneous reconstruction for the number of neighbour devices ranging from 1 to 8 through simulation. 
The 'ntries' column shows the average number of tested share combinations to successfully reconstruct an identifier. 
The computational complexity remains low for high values of $n$ in combination with low number of $k$. 
However, increasing the number of shares beyond $n$=7 in combination with high number of neighbours results in a high load. 
For this reason,  measuring long contact durations should be done by increasing the timeslot duration rather than $n$. 
\begin{table}[!t]
\tiny
\caption{Simultaneous reconstruction performance.}
\begin{center}
\begin{tabular}{ |c | c | c| c | c| c | } 
 \hline
 $k$, $n$:nodes & ntries & $k$, $n$:nodes  	& ntries &  $k$, $n$:nodes &  ntries  \\ 
 \hline
(2, 5), nodes=1	 		&	1	&   (3, 5), nodes=1	& 1			& (4, 5), nodes=1 		&	1\\ 
(2, 5), nodes=2			&	1	&   (3, 5), nodes=2	& 3			& (4, 5), nodes=2 		&	5\\ 
(2, 5), nodes=3	 		&	2	&   (3, 5), nodes=3	& 7			& (4, 5), nodes=3 		&	13\\ 
(2, 5), nodes=4 	 	&	3	&  (3, 5), nodes=4	& 12 			& (4, 5), nodes=4		&	29\\ 
(2, 5), nodes=5 	 	&	4	&  (3, 5), nodes=5	& 19			& (4, 5), nodes=5 		&	56\\ 
(2, 5), nodes=6 	 	&	5	&  (3, 5), nodes=6	& 28			& (4, 5), nodes=6 		&	92\\ 
(2, 5), nodes=7 	 	&	6	&  (3, 5), nodes=7	& 37			& (4, 5), nodes=7		&	144\\ 
(2, 5, nodes=8 	 		&	7	&  (3, 5), nodes=8	& 49			& (4, 5), nodes=8		&	205\\ 
\hline
(3, 6), nodes=1	 		&	1	&   (4, 6), nodes=1	& 	1		& (5, 6), nodes=1 		&	1\\ 
(3, 6), nodes=2			&	3	&   (4, 6), nodes=2	& 	5		& (5, 6), nodes=2 		&	9\\ 
(3, 6), nodes=3	 		&	7	&   (4, 6), nodes=3	& 	16		& (5, 6), nodes=3 		&	37\\ 
(3, 6), nodes=4 	 	&	13	&  (4, 6), nodes=4	&  	35		& (5, 6), nodes=4		&	96\\ 
(3, 6), nodes=5 	 	&	21	&  (4, 6), nodes=5	& 	67		& (5, 6), nodes=5 		&	227\\ 
(3, 6), nodes=6 	 	&	30	&  (4, 6), nodes=6	& 	110		& (5, 6), nodes=6 		&	439	\\ 
(3, 6), nodes=7 	 	&	41	&  (4, 6), nodes=7	& 	176		& (5, 6), nodes=7		&	812\\ 
(3, 6), nodes=8 	 	&	55	&  (4, 6), nodes=8	& 	260		& (5, 6), nodes=8		&	1553\\ 
\hline
(4, 7), nodes=1	 		&	1	&   (5, 7), nodes=1	& 	1		& (6, 7), nodes=1 		&	1\\ 
(4, 7), nodes=2			&	5	&   (5, 7), nodes=2	& 	9		& (6, 7), nodes=2 		&	16\\ 
(4, 7), nodes=3	 		&	18	&   (5, 7), nodes=3	& 	40		& (6, 7), nodes=3 		&	96\\ 
(4, 7), nodes=4 	 	&	42	&  (5, 7), nodes=4	&  	110		& (6, 7), nodes=4		&	342\\ 
(4, 7), nodes=5 	 	&	79	&  (5, 7), nodes=5	& 	256		& (6, 7), nodes=5 		&	1020\\ 
(4, 7), nodes=6 	 	&	131	&  (5, 7), nodes=6	& 	505		& (6, 7), nodes=6 		&	2743\\ 
(4, 7), nodes=7 	 	&	220	&  (5, 7), nodes=7	& 	946		& (6, 7), nodes=7		&	10964\\ 
(4, 7), nodes=8 	 	&	330	&  (5, 7), nodes=8	& 	1961		& (6, 7), nodes=8		&	28735\\ 
\hline
\end{tabular}
\end{center}
\label{tab:simulReconstruction}
\end{table}
It should be noted the results present a worst-case reconstruction performance, which assumes that subsequent shares from the same neighbour cannot be linked together. 
Even if BLE devices use private random address, subsequent shares are likely to be sent using the same private address, which should reduce the actual number tries to reconstruct the device identifier. 
\begin{figure}[]
\centering
\includegraphics[width=2.3in]{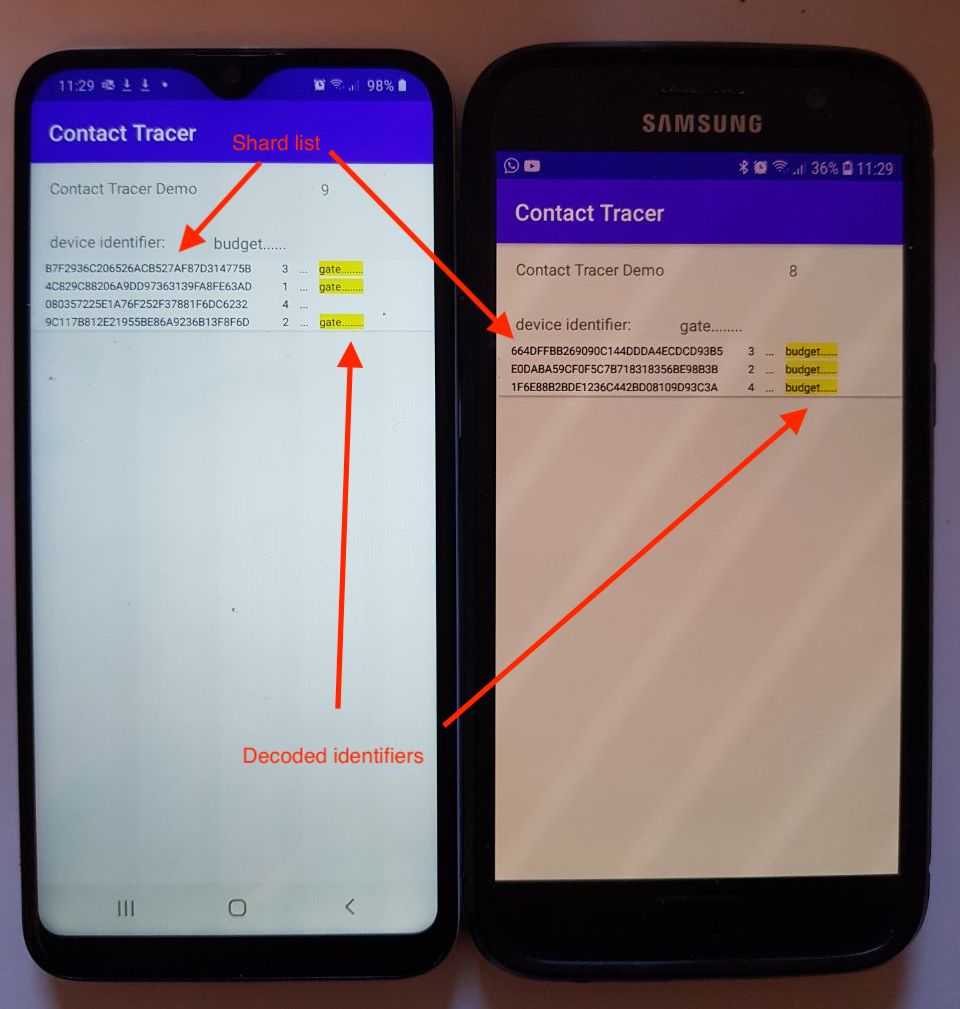}
\caption{Android Implementation running on 2 phones that are transmitting shares and successfully reconstruct each other's identifiers using (k=3, n = 4) scheme.}
\label{fig:implementation}
\end{figure}

\subsection{Android Implementation}
To demonstrate that the proposed approach is feasible on modern smartphones, we have implemented a contact tracing system on Android and tested on  Samsung Galaxy S7 and Galaxy A20e phones. 
The implementation is based on open-source AltBeacon library \cite{altbeacon} and  codahale's Shamir Shared Secret Java implementation \cite{codahale}. 
Table \ref{tab:reconstruction} shows the identifier reconstruction performance on Samsung S7 phone in terms of number of reconstructions per second (rps). 
Each reconstruction cycle includes generating a password string, padding it to 12 bytes, creating a CRC32 checksum and then splitting into $n$ shares. The password string is then reconstructed using $k$ randomly selected shares and the string is verified with the original password string. 
\begin{table}[t]
\tiny
\caption{Android ID reconstruction performance.}
\begin{center}
\begin{tabular}{ |c | c | c| c | c| c |} 
 \hline
 $k$, $n$  		&  	performance, rps & $k$, $n$  	&  	performance  &  $k$, $n$  		&  	performance, rps\\ 
 \hline
(2,5) 	 	& 2906&  	(3,6)&  2777	&(4,7)& 2785\\ 
(3,5) 	 	&2710& 	(4,6)& 	2659		&(5,7)&2450\\ 
(4,5) 	 	&2659& 	 (5,6)& 2538	&(6,7)&2518\\ 
 \hline
\end{tabular}
\end{center}
\label{tab:reconstruction}
\end{table}
\vspace{-1mm}
The results indicate that the performance depends  on both the number of shares $n$ and the minimum number of shares required for reconstruction $k$.  
For example, ($k$=2, $n$=5) configuration  results in 2906 reconstructions per second (rps).
A combination of high $k$  and $n$ ($k$=6, $n$=7) results in the minimum rps of 2518, which can still be computed very fast on modern smartphones.

\subsection{Discussion}

 The implementation shows that the computational complexity depends on the number of shares and maximum number of supported neighbours. 
Both simulation and implementation demonstrate the approach is  feasible to implement on modern smartphones and can drastically reduce the spatio-temporal exposure of users. 
The results show that increasing the number of shares beyond 7 in combination with high number of nodes results in high computational complexity. 
While such configurations may not be desirable for contact tracing application, the high computational complexity could be useful to limit tracking to users with more computational resources only, in what we call 'proof-of-contact' applications. 
The  application of the approach to location based marketing apps, where a user needs to be near a certain location for a certain minimum duration to receive certain ads, can also be an interesting area to explore. 

It is also worth mentioning that a recent update to CDC guidelines contains a clarification on what a "close contact" is. 
The update defines close contact as a cumulative total of 15 minutes over 24 hour duration, whereas the previous guidance defined a close contact as 15 consecutive minutes within six feet of an infected individual. 
The  approach proposed in this paper does not compute a cumulative total, but enforces a minimum contact duration of a single encounter. 
Designing a similar approach to compute a cumulative daily exposure is a potential future work.

\section{Conclusion}

Contact tracing protocols that rely on continuous and periodic beacon transmission, present a serious privacy concern and potentially expose users to promiscuous tracking by malicious users. 
The paper describes a method to limit identity exposure by enforcing a minimum contact duration required for any neighbour node to decode the node identity.  
The method is based on splitting the identifier into multiple shares transmitted separately at random intervals, which requires any scanner to spend a certain duration in the node's neighbourhood to reconstruct the identifier. 
The shares are encoded using Shamir Secret sharing cryptographic algorithm, so that an identity can be reconstructed even if one or more shares are lost due to wireless interference or collision. 

The data-driven evaluation using real BLE dataset has shown that the approach can reduce the overall spatio-temporal exposure from 2.88 to 180 times depending on the secret share algorithm configuration. 
The computational complexity of simultaneous identifier reconstruction from multiple devices can be a limiting factor to the maximum network density and secret share configuration. 
However, the evaluation has shown the ability of method to simultaneously reconstruct up to 7 device identifiers with a reasonable overhead and can be implemented on modern Android phones. 
Applying the idea to reveal node identity depending on the cumulative contact exposure over a day rather than individual contact duration is potential future work.

\bibliographystyle{ieeetr}
\tiny
\bibliography{references}

\begin{thebibliography}{10}

\bibitem{gvili2020contact}
Y.~Gvili, ``{Security Analysis of the COVID-19 Contact Tracing Specifications
  by Apple Inc. and Google Inc.}.'' Cryptology ePrint Archive, Report 2020/428.
\newblock {https://eprint.iacr.org/2020/428}.

\bibitem{cdc}
{Center for Disease Control}, {\em {Public Health Guidance for
  Community-Related Exposure. Accessed 7 Dec 2020}}.

\bibitem{dp3t}
{\em {DP3T Whitepaper}}.
\newblock Available at "https://github.com/DP-3T/documents/blob/master/DP3T
  White Paper.pdf". Accessed 3 December, 2020.

\bibitem{ali2021icoin}
J.~Ali and V.~Dyo, ``{Cross Hashing: Anonymizing encounters in Decentralised
  Contact Tracing Protocols.},'' in {\em {The 35th International Conference on
  Information Networking (ICOIN)}}, IEEE, 13-16 January, 2021.

\bibitem{gaen}
{\em {Apple \& Google Privacy-Preserving Contact Tracing.
  Bluetooth/Cryptography specifications}}.
\newblock Available at \url{https://covid19.apple.com/contacttracing}. Accessed
  3 December, 2020.

\bibitem{nhscovid}
{\em {NHS COVID-19 app}}.
\newblock Available at \url{https://covid19.nhs.uk/}. Accessed 3 December,
  2020.

\bibitem{bluetrace}
{Jason Bay and Joel Kek and Alvin Tan and Chai Sheng Hau and Lai Yongquan and
  Janice Tan and Tang Anh Quy}, ``{BlueTrace. A privacy-preserving protocol for
  community-driven contact tracing across borders. Technical report}.''

\bibitem{cunche2020insa}
M.~Cunche, A.~Boutet, C.~Castelluccia, C.~Lauradoux, D.~L. Métayer, and
  V.~Roca, ``{On using Bluetooth-Low-Energy for contact tracing}.'' Research
  Report. Inria Grenoble Rhône-Alpes; INSA de Lyon. 2020.

\bibitem{pact}
{\em {The PACT Protocol Specification}}.
\newblock Available at \url{https://pact.mit.edu/wp- c
  ontent/uploads/2020/04/The-PACT-protoc ol- specification-ver-0.1.pdf}.
  Accessed 3 December, 2020.

\bibitem{pact2}
{\em {PACT: Private Automated Contact Tracing}}.
\newblock Available at \url{https://pact.mit.edu/}. Accessed 3 December, 2020.

\bibitem{hatke2020using}
G.~F. Hatke, M.~Montanari, S.~Appadwedula, M.~Wentz, J.~Meklenburg, L.~Ivers,
  J.~Watson, and P.~Fiore, ``Using bluetooth low energy (ble) signal strength
  estimation to facilitate contact tracing for covid-19,'' 2020.

\bibitem{Shamir1979acm}
A.~Shamir, ``{How to Share a Secret},'' {\em Commun. ACM}, vol.~22,
  p.~612–613, Nov. 1979.

\bibitem{Boneh2018aicr}
D.~Boneh, M.~Drijvers, and G.~Neven, ``Compact multi-signatures for smaller
  blockchains,'' in {\em Advances in Cryptology -- ASIACRYPT 2018} (T.~Peyrin
  and S.~Galbraith, eds.), (Cham), pp.~435--464, Springer International
  Publishing, 2018.

\bibitem{altbeacon}
{\em {AltBeacon. The Open and Interoperable Proximity Beacon Specification}}.
\newblock Available at \url{https://altbeacon.org}.

\bibitem{ibeacon}
{\em {iBeacon specification}}.
\newblock Available at {https://developer.apple.com/ibeacon/}. Accessed 3
  December, 2020.

\bibitem{eddystone}
{\em {Eddystone specification}}.
\newblock Available at \url{https://developers.google.com/beacons/eddystone}.
  Accessed 3 December, 2020.

\bibitem{ble5}
{Bluetooth SIG}, {\em {Bluetooth Core Specification Version 5.2}}.

\bibitem{sikeridis2018blebeacon}
D.~Sikeridis, I.~Papapanagiotou, and M.~Devetsikiotis, ``{BLEBeacon: A
  Real-Subject Trial Dataset from Mobile Bluetooth Low Energy Beacons},'' 2018.

\bibitem{ble}
{\em {Bluetooth Low Energy}}, Accessed 3 December, 2020.
\newblock Available at
  \url{https://en.wikipedia.org/wiki/Bluetooth_Low_Energy}.

\bibitem{codahale}
{\em {Shamir Secret Sharing}}.
\newblock Available at \url{https://github.com/codahale/shamir}. Accessed 3
  December, 2020.

\end{thebibliography}

\end{document}